# Phonon scattering and stability of $Na_{0.5}CoO_2$

X.N. Zhang [a] P. Lemmens* [a] V. Gnezdilov [b] K.Y. Choi [c] B. Keimer [a] D.P. Chen [a] C.T. Lin [a] F.C. Chou [d]

[a] *Max Planck Institute for Solid State Research, D-70569 Stuttgart, Germany*
[b] *B. I. Verkin Inst. for Low Temperature Physics, NASU, 61164 Kharkov, Ukraine*
[c] *Institute for Materials Research, Tohoku University, Katahira 2-1-1, Sendai 980-8577, Japan*
[d] *Center for Materials Science and Engineering, MIT, Cambridge, MA 02139, USA*

**Abstract**

Raman scattering experiments have been performed on $Na_xCoO_2$ as function of temperature at the composition $x$=0.5 where a structural instability and a metal-insulator transition have been observed. Three additional phonon modes are observed compared to samples with larger $x$. An in-situ annealing study ($T_{max}$ = 550 K) with the initial presence of water vapor at ambient conditions reveals an irreversible structural instability of this composition.

The compound $Na_xCoO_2$ and its hydrated variant $Na_xCoO_2 \cdot yH_2O$ are layered transition metal oxides with several ordering phenomena as function of temperature and composition. For $x\approx1/3$ and $y\approx4/3$ they realize the first known superconducting cobaltate [1]. For $x\geq0.75$ and $y$=0 long range Neel ordering is observed [2]. The Na composition determines the ratio of $Co^{4+}/Co^{3+} = 1-x$ and thereby doping and details of the Fermi surface. The amount of water intercalated in the structure, $y$, affects the $c$ axis lattice parameter and the Na defect structure [3–5].

Recently, further instabilities have been observed for non-hydrated (y=0) samples that show up in transport, magnetic susceptibility and electron diffraction [6–8]. Such effects could arise from strong electronic correlations in the $CoO_2$ layers and/or from order/disorder processes on the partially occupied Na sites. While the physics and chemistry of these anomalies is far from being understood, it seems that the composition with $x$=0.5/$y$=0 is especially sensitive to such instabilities. On the other hand, this composition ($x$=0.5), that shows a metal-insulator transition at $T_{MIT}\approx50$ K and a structural instability at $T_c\approx80$ K, limits the superconducting phase space to higher $x$ [4]. In our Raman study we address this relation between structure and electronic properties of $Na_{0.5}CoO_2$ and investigate its stability with respect to hydration at ambient conditions.

Raman scattering experiments ($\lambda$=514 nm, 6 mW, Ø=100$\mu$m diameter laser focus) have been performed in quasi-backscattering geometry on the $ab$ surface of freshly cleaved single crystals. Samples have been prepared in an optical travelling-solvent floating-zone (TSFZ) furnace [4,8]. After the cleavage the samples were immediately cooled down in He contact gas to prevent any surface degradation. Respective results are shown in Fig. 1.

Two maxima corresponding to the $E_{1g}$ and the $A_{1g}$ eigenmodes of the $CoO_6$ octahedron are observed with frequencies at 570 and 475 $cm^{-1}$, very similar to those frequencies observed in single crystals with higher Na content, x=0.83 (580.7 and 475 $cm^{-1}$) [9]. The smaller frequency of the out-of-plane $A_{1g}$ mode is pointing towards a weaker interlayer coupling for the sample with smaller Na content. Three additional modes (414, 437 and 497 $cm^{-1}$) exist for x=0.5. The latter mode has the weakest intensity. In addition there exist a constant background of scattering attributed to electronic scattering [9] with a cut-off frequency of about 500 $cm^{-1}$. As function of temperatures we observe only a gradual increase of the intensities of the phonons. The normalized linewidth shows a non-monotonous temperature dependence with a minimum between 60-100 K (not shown here). We conclude a moderate coupling of the oxygen modes within the $CoO_2$ layers to the electronic degrees of freedom.

* MPI-FKF, D-70569 Stuttgart, Heisenbergstr. 1, phone: +49 (0)711 6 89 - 17 54, fax: +49 (0)711 6 89 - 16 32, email: p.lemmens@fkf.mpg.de

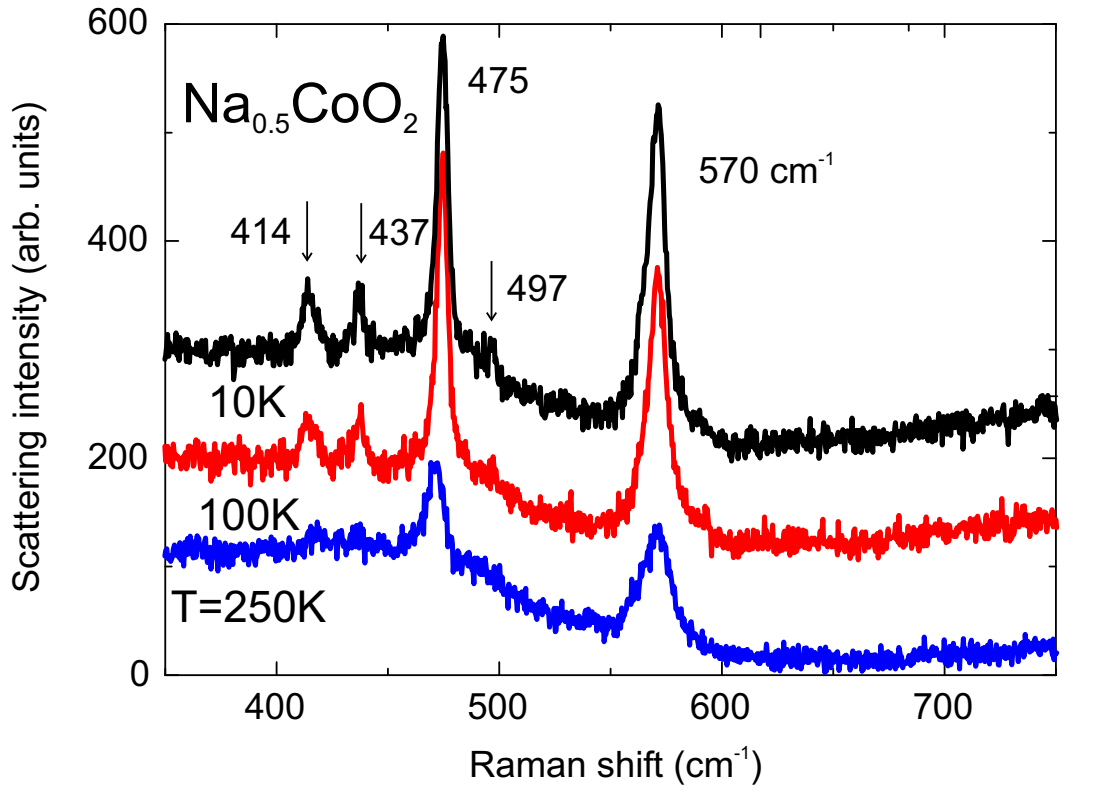

Fig. 1. Raman scattering spectra of $Na_{0.5}CoO_2$ with in-plane light polarization at low temperatures.

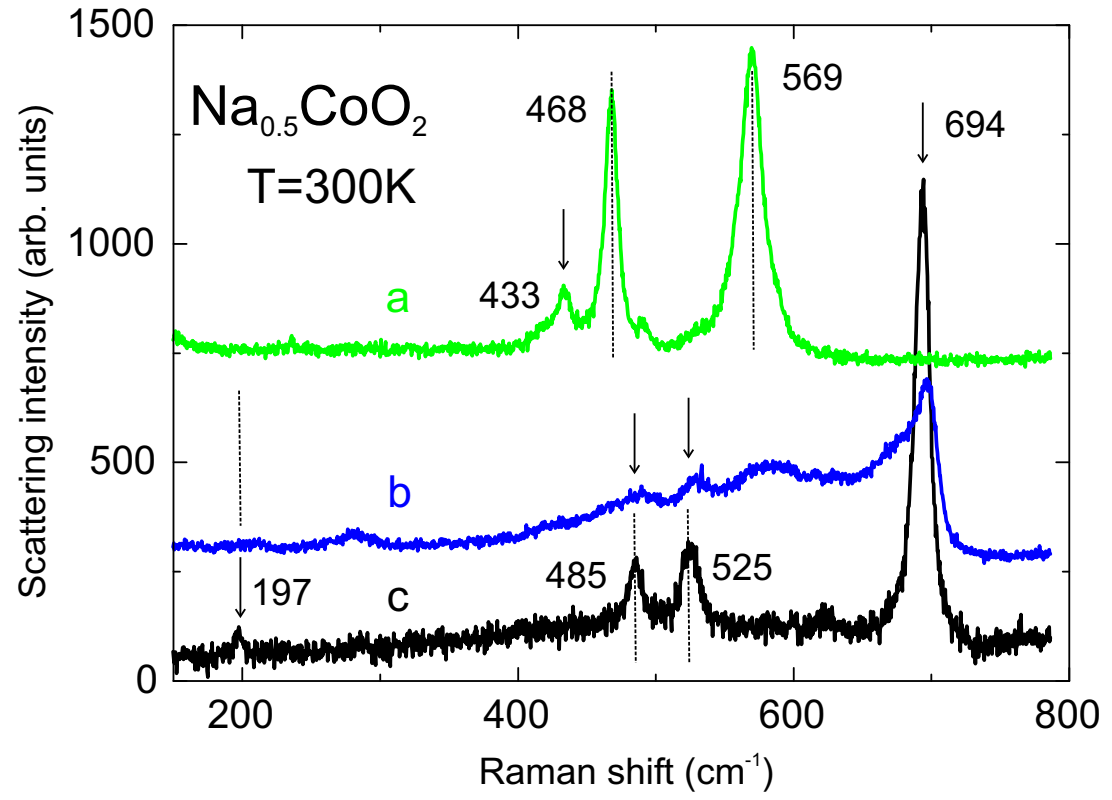

Fig. 2. Raman scattering spectra at T=RT a) before and b) after an exposure to humidity at ambient conditions and c) after subsequent annealing.

The sample has then been warmed up and exposed to water vapor at room temperature (RT). Fig. 2 shows three spectra, a) following the water exposure, b) after a waiting time of 24 h, and c) after annealing the sample on an in-situ heating stage with $T_{max} = 550$ K. The initial exposure to water vapor leads to an enhancement of the main phonon lines. Similar effects exist for the superconducting, hydrated $Na_xCoO_2 \cdot 1.3H_2O$ with x=0.3 [9]. These observations are attributed to a smaller effective disorder on the Na site with decreasing interlayer coupling.

However, in contrast to the superconducting phase, $Na_{0.5}CoO_2$ is unstable with respect to hydration as shown in spectra b)+c). After some time at RT and especially after an annealing treatment phonons with different frequencies exist that cannot be attributed to the original spectrum of freshly cleaved single crystals. It is interesting to note, that similar spectra as shown in Fig. 2c) have been reported for "cleaned" crystals of $Na_xCoO_2$ grown from a NaCl flux [10]. Degradation effects and time dependencies have also been reported earlier on such crystals [11]. It is possible that solving the crystals from the bulk of the NaCl flux using $H_2O$ leads to a Na inhomogeneity and a variation of doping at the surface. The instability of the composition $Na_{0.5}CoO_2$ with respect to hydration leads probably to the observed degradation processes.

The authors acknowledge fruitful discussions with G. Khalliulin, C. Bernhard, R. Kremer, and Yu. Pashkevich. This work was supported by the MRSEC Program of the National Science Foundation under award number DMR 02-13282, the DFG SPP1073, and INTAS 01-278.